\documentstyle[12pt,epsfig]{article}

\newcommand{\be}{\begin{eqnarray}}
\newcommand{\ee}{\end{eqnarray}}

\def\anue{{\bar\nu_e}}

\newcommand{\dm}{\mbox{$\Delta m_{21}^2$~}}

\newcommand{\kl}{\mbox{KamLAND~}}

\def\ltap{\ \raisebox{-.4ex}{\rlap{$\sim$}} \raisebox{.4ex}{$<$}\ }

\newcommand{\sss}{\sin^2 \theta_{12}}
\newcommand{\ms}{\Delta m^2_{21}}
\begin{document}
%%%%%%%%%%%%%%%%%%%

\begin{flushright}
SISSA 47/2004/EP     \\
%SINP/TNP/03-xx\\
hep-ph/0406328
\end{flushright}

\begin{center}
{\Large \bf Update of the Solar Neutrino Oscillation Analysis   
with the 766 Ty \kl Spectrum
}
\vspace{.5in}

{ Abhijit Bandyopadhyay$^1$,
%\footnote{e-mail: abhi@theory.saha.ernet.in},
Sandhya Choubey$^{2,3}$,
%\footnote{email: sandhya@he.sissa.it},
Srubabati Goswami$^{4}$,
%\footnote{e-mail: sruba@mri.ernet.in},
S.T. Petcov$^{3,2,5}$,
%\footnote{e-mail: petcov@he.sissa.it}
D.P. Roy$^{6,7}$}
%\footnote{e-mail:dproy@tifr.res.in}}
\vskip .5cm

$^1${\small {\it Saha Institute of Nuclear Physics}},
{\small {\it 1/AF, Bidhannagar,
Calcutta 700 064, India}},\\

$^2${\small {\it INFN, Sezione di Trieste, Trieste, Italy}},\\
$^3${\small {\it Scuola Internazionale Superiore di Studi Avanzati,
I-34014,
Trieste, Italy}},\\
$^4${\small {\it Harish-Chandra Research Institute, Chhatnag Road, Jhusi,
Allahabad  211 019, India}},\\
$^5${\small {\it Institute of Nuclear Research and Nuclear Energy, 
Bulgarian Academi of Science, Sofia, Bulgaria}},\\
$^6${\small{\it The Abdus Salam International Centre for Theoretical Physics,
I-34100, Trieste, Italy}},\\
$^7${\small {\it Tata Institute of Fundamental Research, Homi Bhabha Road, 
Mumbai 400 005, India}}

\vskip 1in

\end{center}
\begin{abstract}
We investigate the impact of 
the 766.3 Ty KamLAND spectrum data on the determination 
of the solar neutrino oscillation parameters.
We show that the observed spectrum distortion 
in the \kl experiment 
firmly establishes
\dm to lie in the low-LMA solution region.
The high-LMA solution is 
excluded at more than  4$\sigma$ 
by the global solar neutrino and \kl spectrum data.
The maximal solar neutrino mixing is 
ruled out at $6\sigma$ level.
The $3\sigma$ allowed region in the 
$\ms-\sss$ plane is found to be  
remarkably stable with respect to 
leaving out the data from one 
of the solar neutrino experiments 
from the global analysis.
We perform a three flavor
neutrino oscillation analysis of the global solar neutrino 
and \kl spectrum data as well.
The $3\sigma$ upper limit on $\sin^2\theta_{13}$ 
is found to be $ \sin^2\theta_{13} <0.055$.
We derive predictions 
for the CC to NC event rate ratio
and day-night (D-N) asymmetry in the CC event rate,
measured in the SNO experiment, and for the 
suppression of the event rate in the BOREXINO 
and LowNu experiments.
Prospective high precision
measurements of the solar neutrino 
oscillation parameters are also discussed.

\end{abstract}

\newpage

%%%%%%%%%%%%%%%%%%%%%%
\section{Introduction}
\vspace{-0.3cm}
%%%%%%%%%%%%%%%%%%%%%%

  The last four years  will most likely be
described in the future as the golden years of solar neutrino physics. 
The pioneering results of the Homestake experiment \cite{cl},
which first observed neutrinos emitted by the Sun and 
discovered the existence of a solar neutrino deficit
\footnote{Let us recall that the Cl-Ar method of 
neutrino detection, used in the Homestake experiment, 
was first proposed in \cite{BPont46}.},
and of the Kamiokande \cite{kam}, 
SAGE and GALLEX/GNO \cite{ga} experiments,
which confirmed and extended the Homestake results
on the solar neutrino deficit,  
were reinforced during the last four years 
by a series of precision measurements by 
the Super-Kamiokande(SK) \cite{sk}, SNO  
\cite{Ahmad:2001an},\cite{ Ahmad:2002jz},\cite{ Ahmed:2003kj}
and \kl \cite{kl162} experiments. With the recent 
publication of the \kl 766 Ty spectrum data \cite{kl766}
and under the plausible assumption of CPT-symmetry,
for the first time a unique solution
of the solar neutrino problem in terms of 
neutrino oscillations \cite{BPont5767,msw} can be
unambiguously identified. Let us summarize the main 
steps of the progress in our understanding of 
the solution of the solar neutrino problem, 
made in this past four years.

\begin{itemize}
\item The first charged current (CC) data from SNO \cite{Ahmad:2001an}
together with absence of distortions 
of the spectrum of the final state $e^-$
in the $\nu - e^-$ elastic scattering 
reaction due to solar neutrinos, measured with a 
high precision in the SK experiment  
\cite{sk}, excluded the 
vacuum oscillation (VO) and the small 
mixing angle (SMA) MSW \cite{msw} 
solutions in favour of the large mixing angle 
(LMA) MSW and LOW solutions \cite{snocc,snoccothers}.
\item The first neutral current (NC) solar neutrino data  from SNO 
\cite{ Ahmad:2002jz},
obtained by observing the solar neutrino capture on $D_2O$ 
(Phase I of the experiment), provided a direct 
estimate of the Boron neutrino flux normalisation $f_B$. It confirmed the 
Standard Solar Model (SSM) prediction for this quantity
(the uncertainty in the experimentally determined $f_B$
was smaller than that in the SSM prediction).
This implied that the CC rates of Cl, SK 
and SNO are indeed smaller than 0.5. 
This strongly favoured the LMA MSW solution over the LOW solution 
and a non-maximal solar neutrino mixing angle 
\cite{snonc}. 
\item The convincing evidence in favour of the LMA solution 
was obtained in the $\kl$
experiment with reactor $\bar{\nu}_e$ \cite{kl162},
which published first results, 
based on statistics of 162 Ty,
in December of 2002.
Under the plausible assumption of CPT invariance,  
the suppression of the reactor $\bar{\nu}_e$ flux
observed in the \kl experiment firmly established the 
LMA solution, ruling out the LOW solution at about 5$\sigma$ level.
Moreover, the 162 Ty \kl data on the $e^+-$spectrum distortion,  
combined with the global solar neutrino data,
implied that the LMA solution 
was confined to two sub-regions :  
low-LMA (or LMA-I), centered around $\Delta m^2 = 7.2 \times 10^{-5} 
$ eV$^2$, and high-LMA (or LMA-II) with $\Delta m^2$ 
centered around $\Delta m^2 = 1.5 \times 10^{-4}$ eV$^2$. 
It was found that in both cases
$\sin^2\theta \approx 0.3 $. 
The best-fit  was in the 
low-LMA region, while the high-LMA region 
was allowed only at 99\% C.L. 
\cite{solfit1,solfit2}
\item Finally the NC data from the salt phase of the SNO experiment
\cite{Ahmed:2003kj},
provided a more precise measurement of $f_B$. 
The inclusion of these
data in the global solar neutrino oscillation analysis
reduced further the allowed region 
of solar neutrino 
oscillation parameters.
Now the high-LMA region was allowed 
only at 2.65 $\sigma$ level, while the maximal solar neutrino 
mixing was excluded 
at about 5$\sigma$ \cite{snosaltus, snosaltothers}.
\end{itemize}

   One of the most important issues 
after these developments was the 
definitive resolution of the high-LMA
and low-LMA solution ambiguity. 
This was expected to lead to a precise 
determination of the neutrino mass squared
difference driving the solar neutrino oscillations. 
For this reason a study of \kl  410 Ty and 1000 Ty 
simulated spectrum data, corresponding to different points 
in the parameter space, spanning 
the low-LMA and high-LMA solution regions, 
was made in \cite{snosaltus}.  
This study showed, in particular, that 
if the true \kl 1000 Ty spectrum data
corresponded to a point in the low-LMA region, 
the high-LMA solution would be ruled out at 3$\sigma$ level 
by the combined solar and \kl data, while
the low-LMA solution region will be considerably reduced.
The best-fit point obtained from the analysis of the 
energy spectrum of the recently released
766.3 Ty data from \kl indeed lies 
inside the low-LMA region
\cite{kl766}. And 
the combined global solar and \kl data indeed excludes 
the high-LMA solution at 3$\sigma$ level, 
in agreement with the above expectation. 

   In this article we investigate the impact of 
the 766.3 Ty KamLAND spectrum data on the determination 
of the solar neutrino oscillation parameters.
We perform first a two-neutrino oscillation analysis of the global 
solar neutrino and the latest KamLAND spectrum data.
This permits us to quantify the improvements 
in the precision of determination of the 
solar neutrino oscillation parameters
which the new KamLAND data imply
and, in particular, to assess the status
of the high-LMA solution. We check the 
stability of the allowed region of values 
of the solar neutrino oscillation parameters thus derived
with respect to leaving out 
from the analysis the data
from one of the solar neutrino experiments.
This serves also as a check of the consistency 
between the data from the different experiments 
and gives some idea about the level of redundancy of 
the global solar neutrino data set. We next extend
the analysis to the case of three neutrino oscillations.
We derive, in particular, a new upper limit on
the CHOOZ mixing angle $\theta_{13}$, and study
the dependence of the allowed values of the parameters
$\Delta m^2_{21}$ and $\sin^2\theta_{12}$
which drive the solar neutrino oscillations,
on the value of $\sin^2\theta_{13}$.
We give predictions 
for the CC to NC event rate ratio
and day-night (D-N) asymmetry in the CC event rate,
measured in the SNO experiment, for the 
suppression of the event rate in the BOREXINO 
and LowNu experiments, designed to measure the
$^7Be$ and $pp$ solar neutrino fluxes.
Finally, we discuss also how the precision of $\sin^2\theta_{12}$
determination can improve with the increasing of the precision of the
future SNO data, as well as, by performing a reactor 
$\bar{\nu}_e$ oscillation experiment with a baseline
$\sim 70$ km.
 
\section{Two Flavour-Neutrino Oscillation Analysis}

\begin{figure}[t]
\centerline{\psfig{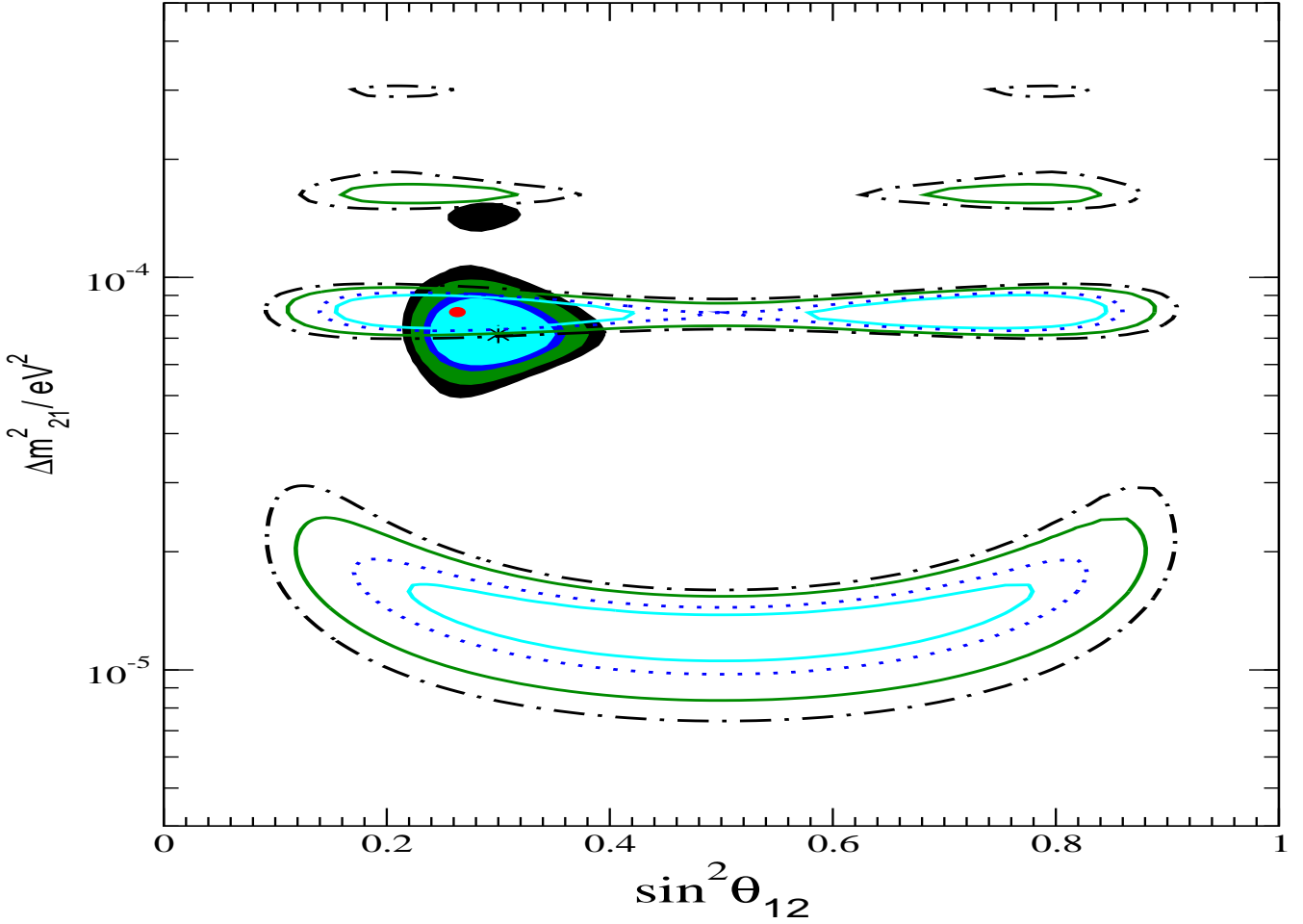}}
\caption{\label{klspec}
The 90\%, 95\%, 99\% and 99.73\% C.L. 
allowed regions in the $\dm-\sss$ plane from a 
$\chi^2$-analysis i) of the  766.3 Ty $\kl$ data 
(white areas within contour lines),
and ii) of the global solar 
neutrino and 162 Ty $\kl$ data (shaded areas). 
The best fit points in the two cases are marked by
a black dot and by a star, respectively.
}
\end{figure}

We first present the results of a standard 
two-flavor neutrino oscillation analysis.
We use the 13 bin
\kl spectrum data and define a $\chi^2$ 
assuming a Poisson distribution as
%%%%%%%%%%%%%%%%%%%%%%%%%%%%%%%%%%%%%%
\be
\chi^2_{klspec}=
\sum_{i}\left[2(X_n S_{KL,i}^{theory} - S_{KL,i}^{expt}) 
+ 2 S_{KL,i}^{expt} \ln(\frac{S_{KL,i}^{expt}}
{X_n S_{KL,i}^{theory}})\right] + \frac{(X_n -1)^2}{\sigma^2_{sys}},
\label{chip}
\ee
%%%%%%%%%%%%%%%%%%%%%%%%%%%%%%%%%%%%%%%%%
where 
$X_n$ is 
allowed to vary freely and 
$\sigma_{sys}$ is 
the systematic uncertainty.
%taken to be 6.5\%   
\footnote 
{Note that several
theoretical and experimental systematic errors
like those due to energy scale, reactor spectrum etc. 
are energy dependent. 
However, these details being inacessibile to us,  we have used 
the same
total systematic  
error
% of 6.5\% 
for all the bins. 
Detailed information by the \kl 
collaboration on the errors and their correlations 
in each bin 
will make our analysis more accurate. 
As \kl statistics is increased and 
systematics start to dominate more detailed information from the \kl 
collaboration will be an important requirement.} 
%and $X_n$ is 
%allowed to vary freely.
 We include the revised 
resolution width, fuel composition, 
detector fiducial mass
and efficiencies from \cite{kl766}. 
The other details of our analysis can be found in \cite{solfit1,prekl}.
Some of the reactors, particularly 
the Kashiwazaki-Kariwa and Fukushima I and II reactor complexes, were 
partially/totally shut-down during some of the period of data taking 
in KamLAND. We have approximately 
taken into account this change in the reactor flux due to the 
reactor shut-down using the plots showing the time variations 
of the number of fissions in a given reactor and 
hence the expected reactor $\anue$ flux in KamLAND \cite{talknove}.
We have also used the information on the reactor operation schedules 
available on the web \cite{web}. 
In the latest version of \cite{kl766}, the \kl collaboration have
identified a new source of background in the their analysis, coming
from $^{13}C(\alpha,n)^{16}O$ reaction induced by the $\alpha$ decay of
the radon daughter $^{210}Po$ in the liquid scintillator. This increases
the total background in their $\anue$ signal to $17.8\pm 7.3$ above
$E_{vis}=2.6$ MeV. We include in our analysis,
this new background and its associated uncertainty.

  We show by the unshaded contours in 
Fig. \ref{klspec} the allowed regions 
obtained using the 766.3 Ty \kl spectrum data. 
The best-fit according to our analysis is at 
\footnote{ 
With the KamLAND results in the earlier versions of
\cite{kl766} which did not include 
the new background, the 
best-fit values were 
$\Delta m^2_{21} = 8.4 \times 10^{-5}$ eV$^2$,
$\sin^2\theta_{12} =0.24$ and 
$\chi^2_{min}/d.o.f. = 19.46/10$.
Thus, the inclusion of the 
new background in our analysis
is seen to have changed the best-fit oscillation parameters  
as well as  improved the goodness of fit (g.o.f). This is in agreement with 
what the \kl collaboration has obtained.}  
%%%%%%%%%%%%%%%%%%%%%%%%%%%%%%
\begin{equation}
% \begin{itemize}
% \item
% $\Delta m^2_{21} = 8.4 \times 10^{-5}$ eV$^2$, 
% $\sin^2\theta_{12} =0.24$ ($\tan^2\theta_{12}=0.32$).
% \end{itemize}
\Delta m^2_{21} = 8.2 \times 10^{-5}~{\rm eV^2},~~~ 
\sin^2\theta_{12} =0.26, ~~~ \chi^2_{min}/d.o.f. = 15.24/10
% ($\tan^2\theta_{12}=0.32$).
\label{eq2}
\end{equation}
%%%%%%%%%%%%%%%%%%%%%%%%%%%%%
The best-fit value of $\Delta m^2_{21}$ we find 
agrees reasonably well with that obtained by the $\kl$ 
collaboration \cite{kl766},
while our best fit value of $\sin^2\theta_{12}$ 
% is a little lower 
is somewhat lower than that found in \cite{kl766} 
because of differences in the fitting procedure
and the relative insensitivity of the KamLAND data
to this parameter.
The regions at $\Delta m^2 \geq 2 \times 10^{-4}$ eV$^2$
which were allowed by the \kl 162 Ty spectrum data \cite{kl162},
are now severely disfavored
due to increased precision on the observed spectral
distortion and only a very tiny area
is allowed at the 3$\sigma$ level.
Superposed on the same figure, we show by the shaded areas the 
allowed regions obtained using   
the combined solar neutrino + 162Ty \kl data. 
As it follows from this figure, 
the best-fit point 
of the new \kl spectrum data lies inside the 
allowed low-LMA region, obtained in the solar neutrino + 162 Ty 
\kl spectrum data analysis. 

%%%%%%%%%%%%%%%%%%%%%%%%%%%%%%%%%%%
\begin{figure}[t]
\centerline{\psfig{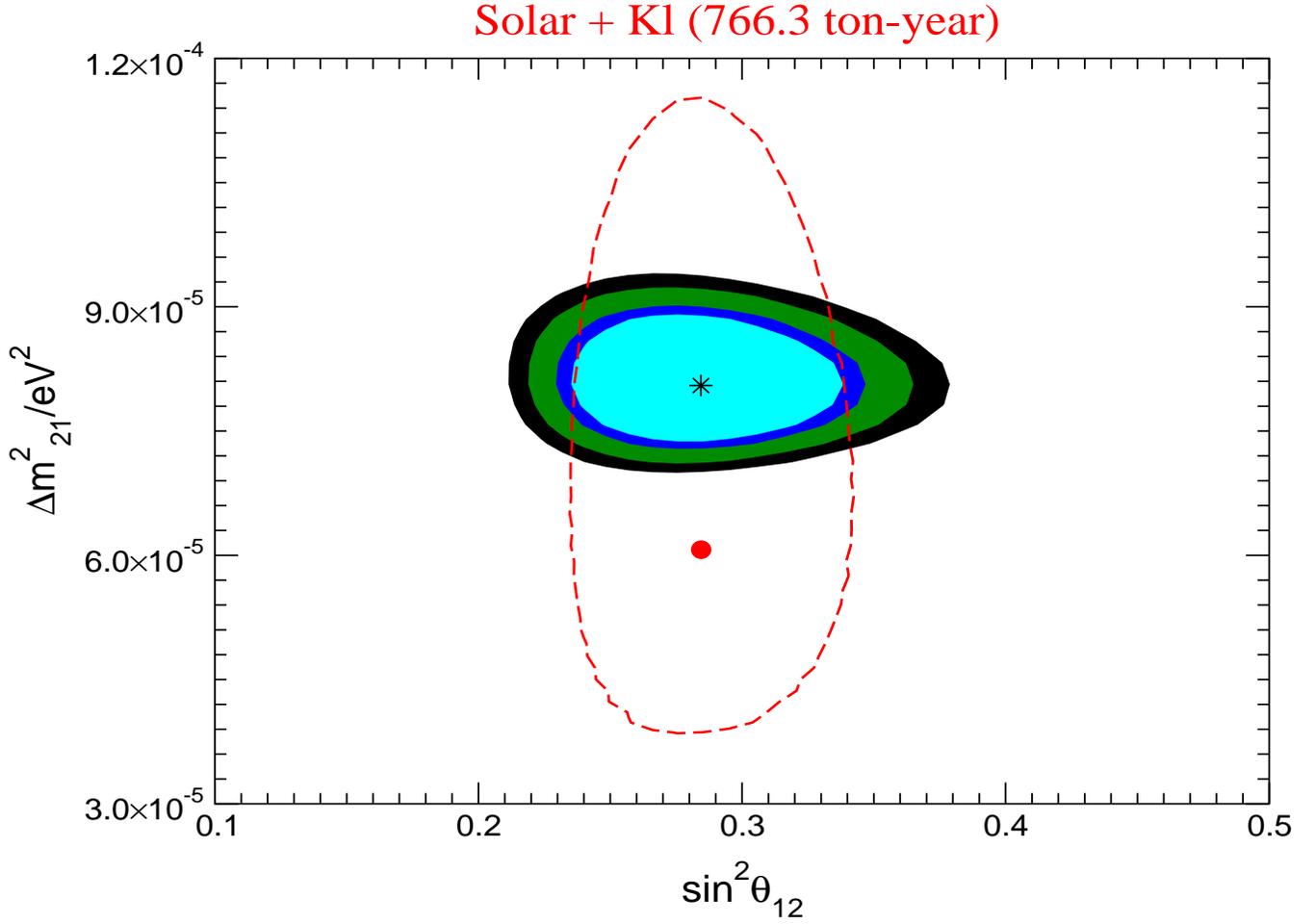}}
\caption{\label{klspec+solar}
The 90\%, 95\%, 99\% and 99.73\% C.L. 
allowed regions in the $\dm-\sss$ plane, 
obtained in a combined $\chi^2$-analysis of the global 
solar neutrino and the 766.3 Ty \kl spectrum data 
(shaded areas).
The region allowed by the solar neutrino data alone 
at 90\% C.L. is also shown 
for comparison (region within the dashed-line contour). 
The best-fit points are marked in both cases.}
\end{figure}
%%%%%%%%%%%%%%%%%%%%%%%%%%%%%

In Fig. \ref{klspec+solar} we show the allowed region obtained from 
the combined analysis of the global 
solar neutrino data and the 766.3 Ty \kl spectrum data.
The dashed line in the figure indicates the region allowed at 
90\% C.L. by the global solar neutrino data alone. The 
star (dot) marks the best-fit point 
of the solar neutrino + 766.3 Ty \kl (solar neutrino) data.
We have used in this analysis 
the solar neutrino data on the total event rates 
from the radiochemical experiments 
Cl \cite{cl} and Ga (Gallex, SAGE and GNO combined) \cite{ga}, 
the 1496 day 44 bin Zenith 
angle spectrum data from SK \cite{sk}, 
the combined CC, NC and Electron Scattering (ES)
34 bin energy spectrum data from the
phase I ( pure $D_2O$ phase) of SNO \cite{Ahmad:2002jz} 
and the data on CC, NC and ES total observed rates
from the phase II (salt phase) of the SNO experiment
\cite{Ahmed:2003kj}.

For the combined analysis of solar and \kl data we define the 
global $\chi^2$ as 
\be 
\chi^2_{global} = \chi^2_{\odot} + \chi^2_{klspec}
\ee
where
\be
\chi^2_{\odot} = \sum_{i,j=1}^N (R_i^{\rm expt}-R_i^{\rm theory})
(\sigma_{ij}^2)^{-1}(R_j^{\rm expt}-R_j^{\rm theory})
\label{chi2}
\ee
where $R_{i}$ are the solar data points, $N$ is
the number of data points and
$(\sigma_{ij}^2)^{-1}$ is the inverse of the covariance matrix,
containing the squares of the correlated and uncorrelated experimental
and theoretical errors.
The $^8B$ flux normalisation factor 
$f_B$ is left to vary freely in the analysis. 
For the other solar neutrino fluxes
($pp$, $pep$, $^7Be$, $CNO$, $hep$),
the predictions and estimated uncertainties 
of the most recent standard solar model (SSM) 
\cite{bp04} (BP04) have been utilized.
For further details of our solar neutrino data 
analysis we refer the reader
to our earlier papers \cite{snocc,snonc,snosaltus}.

\begin{figure}[t]
\centerline{\psfig{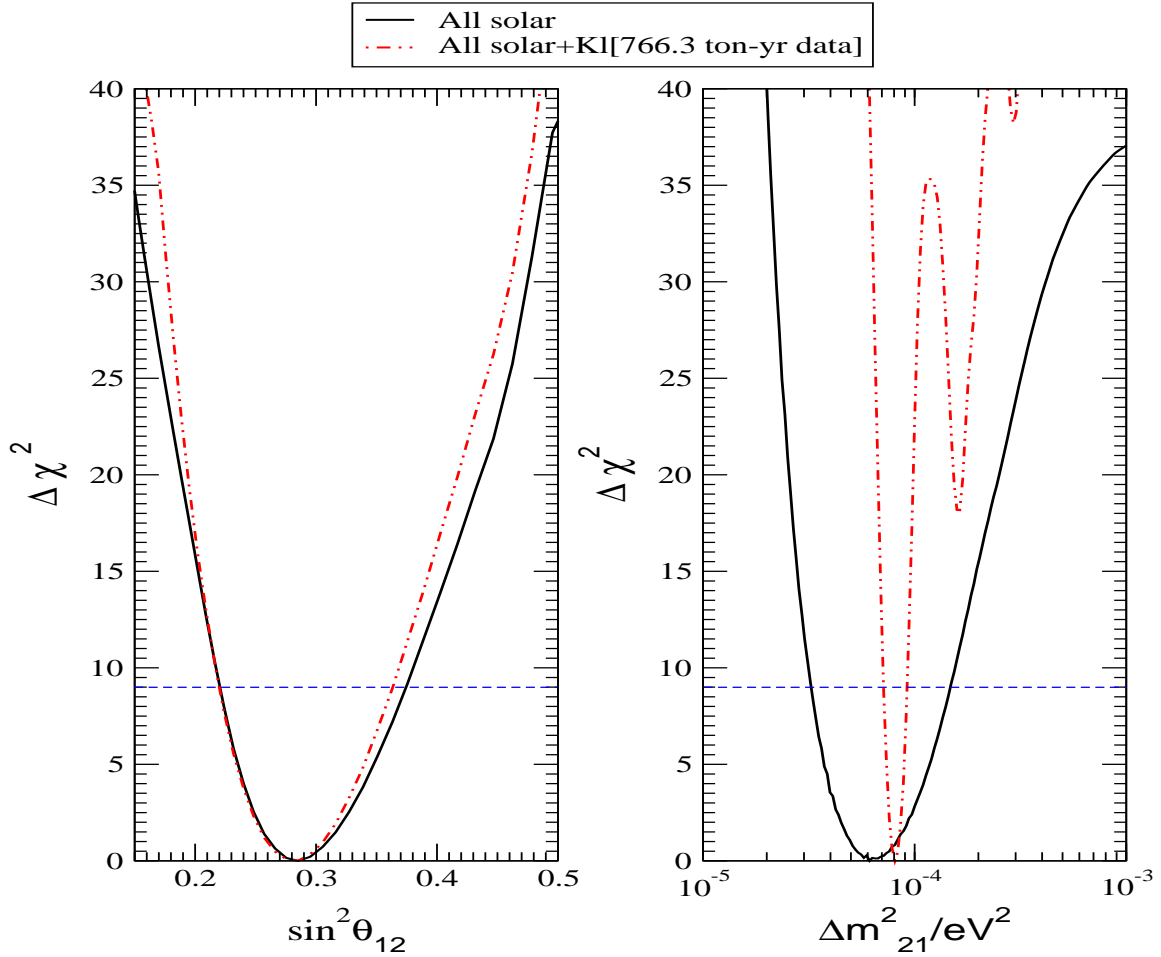}}
\caption{\label{delvs12}
Bounds on \dm and $\sss$ from
the $\Delta \chi^2 $ as a function of \dm and $\sss$, respectively.
The results shown in both panels are obtained by allowing
all the other parameters to vary freely.
The dashed line shows the $3\sigma$ limit corresponding
to 1 parameter fit.
}
\end{figure}

We find that with the inclusion of the latest 
\kl spectrum data, the high-LMA region 
is disfavored at more than 99.9\% C.L. in a 2 parameter fit.
Thus,  the high-LMA solution
is excluded at more than 3$\sigma$. 
This establishes the low-LMA 
solution as 
unique solution of the 
solar neutrino problem. 
It also confirms our prediction 
\cite{snosaltus} that, if the 
best-fit of the \kl  spectrum data
corresponds to a point in the 
low-LMA solution region, 
there will be no high-LMA region 
allowed at 3$\sigma$ level. 
The best-fit point we get from the 
combined solar neutrino and \kl 
data analysis is,
%%%%%%%%%%%%%%%%%%%%%%%%%%
% \begin{itemize}
% \item
% $\Delta m^2_{21} = 8.3 \times 10^{-5}$ eV$^2$, $\sin^2\theta_{12}=0.27$,
% $\mathrm f_B$ = 0.89,
% \end{itemize}
\begin{equation}
\Delta m^2_{21} = 8.0 \times 10^{-5}~{\rm eV^2},~~ \sin^2\theta_{12}=0.28,~~
\mathrm f_B = 0.88, ~~\chi^2_{min}/d.o.f. = 85.42/92
\label{eq3}
\end{equation}
%%%%%%%%%%%%%%%%%%%%%%%%%%
in good agreement with that obtained in \cite{kl766}.  
Note that the best-fit point 
from the global solar neutrino data analysis, 
%%%%%%%%%%%%%%%%%%%%%%%%%%%%%
% \begin{itemize}
% \item
% $\dm$ = 6.06 $\times 10^{-5}$ eV$^2$,
% $\sss$ = 0.3, $\mathrm f_B$ = 0.89, 
% \end{itemize}
\begin{equation}
\dm = 6.07 \times 10^{-5}~{\rm eV^2},~~~
\sss = 0.29,~~~\mathrm f_B = 0.90, ~~~\chi^2_{min}/d.o.f. = 69.06/80
\label{eq4}
\end{equation}
%%%%%%%%%%%%%%%%%%%%%%%%%%%%
lies outside the 3$\sigma$ range, allowed after including the 
new \kl data. However, the $\chi^2$ function 
for the solar neutrino data changes
weakly when $\dm$ varies in an interval of values
which is centered on the best fit value in eq. (\ref{eq4})
and includes the best fit value of the global 
data set, eq. (\ref{eq3}) (see Fig. \ref{delvs12}).
% We note that 
The best-fit value of $\Delta m^2_{21}$ in the global fit is
controlled by the \kl data, 
whereas the best-fit value of $\sin^2\theta_{12}$ 
is controlled by the global solar neutrino data. 
The $\ms$ allowed region is seen to 
have narrowed down considerably making it possible 
to plot it on a linear scale.

%%%%%%%%%%%%%%%%%%%%%%%%%%%%%%%%%%% 
\begin{table}
\begin{center}
\begin{tabular}{ccccc}
\hline
{Data set} & (3$\sigma$)Range of & (3$\sigma$)spread in  & 
(3$\sigma$) Range of
&(3$\sigma$) spread in \cr
{used} & $\Delta m^2_{21}$ eV$^2$
& {$\Delta m^2_{21}$} & $\sin^2\theta_{12}$
& {$\sin^2\theta_{12}$} \cr
\hline
{only sol} & 3.0 - 17.0
&{70\%} & $0.21-0.39$ &30\%\cr
{sol+162 Ty KL}& 4.9 - 10.7 
& 37\%
& $ 0.21-0.39$ & 30\%  \cr
{sol+ 766.3 Ty KL}& 7.0 - 9.4
& 15\% &
$0.21-0.38$ & 29\% \cr
\hline
\end{tabular}
\caption{\label{spread}
3$\sigma$ allowed ranges of $\Delta m^2_{21}$ and
$\sin^2\theta_{12}$
from the analysis of the global solar neutrino, 
and global solar neutrino + \kl (past and present) data.
% analysis of past, present and future.
We show also the \% spread in the 
allowed values of the two neutrino oscillation 
parameters.}
\end{center}
\end{table}
%%%%%%%%%%%%%%%%%%%%%%%%%%%%%%%%%%%%%%%%%%%%%%%%

In  Table \ref{spread} we present the  3$\sigma$ 
allowed ranges of $\Delta m^2_{21}$ 
and $\sin^2\theta_{12}$, obtained  
using different data sets. 
We also show the uncertainty in the value of the 
parameters through a quantity ``spread'' which we define as
%%%%%%%%%%%%%%%%%%%%%%%%%%%%%%%%%%%%%
\be
{\rm spread} = \frac{ prm_{max} - prm_{min}}
{prm_{max} + prm_{min}}\times 100,
\label{error}
\ee
%%%%%%%%%%%%%%%%%%%%%%%%%%%%%%%%%%%%%%
%
where ${prm}$ denotes the parameter \dm or $\sin^2\theta_{12}$,
and $prm_{max}$ and $prm_{min}$ are the maximal and minimal values of
the chosen parameter allowed at a given C.L.
Table 1 illustrates the  
remarkable sensitivity of the \kl 
experiment to $\Delta m^2_{21}$,
which results in stringent constraints
on the allowed values of $\Delta m^2_{21}$.
However, the \kl experiment 
does not constrain the allowed range 
of $\theta_{12}$ much better 
than the solar neutrino experiments.

%%%%%%%%%%%%%%%%%%%%%%%%%%%%%%%%%%%%
%%%%%%%%%%%%%%%%%%%%%%%%%%%%%%%%%%%%%%%%
So far we have given the allowed regions of 
\dm and $\sss$,
obtained from the two parameter 
fits of the data. It is instructive to see 
the bounds on the oscillation 
parameters using one parameter plots of  
$\Delta \chi^2 = \chi^2 - \chi^2_{min}$ vs. $\ms$ and vs. $\sss$.
In Fig. \ref{delvs12} we show the dependence
of $\Delta \chi^2$ on $\sss$ (left hand panel) and on 
$\dm$ (right hand panel) 
after marginalising over the remaining free parameters. 
In this analysis  
only the solar neutrino data, and the solar neutrino 
+ 766.3 Ty \kl spectrum data have been used.
We find that the  allowed  
range of $\Delta m^2_{21}$ values becomes much 
narrower compared to that obtained 
using only the global solar neutrino data.  
The inclusion of the recent \kl results makes the  
$\Delta \chi^2$ for the high-LMA region even larger, 
excluding it at more than 4$\sigma$ for a one parameter fit. 
From this figure we also see that 
the best-fit value of $\dm$, 
obtained from the global solar neutrino data, 
has a $\Delta \chi^2 > 40$,
and hence is disfavored at $>6\sigma$.
The inclusion of the new \kl spectrum data 
disfavors maximal solar neutrino mixing 
to a greater degree. The $\Delta \chi^2$ value at
$\sin^2\theta_{12} = 0.5$
is a little above 40, thereby  
excluding the maximal mixing
at more than 6$\sigma$ for a one parameter fit.  
Figure \ref{delvs12}, showing the dependence of
$\Delta \chi^2$ on $\sin^2\theta_{12}$, corroborates our 
results presented in Table 1, namely,
that the allowed range of $\sin^2\theta_{12}$ does not change 
considerably up to the $3\sigma$ level
with the inclusion of the new \kl results.  
The reason for this can be traced to the fact that 
for the values of $\ms$ and $\sss$ 
allowed by the combined solar neutrino and \kl data, 
the Earth matter effects are negligible 
at the baselines relevant for the \kl experiment 
and the relevant $\bar{\nu}_e$ survival probability 
reads:
%%%%%%%%%%%%%%%%%%%%%%%%%%%%%%%%%%%%%%%%
\be
P_{ee}^{KL}\approx1- \sin^22\theta_{12} 
\sin^2\left(\frac{\pi L}{\lambda}\right),
\label{probkl}
\ee
%%%%%%%%%%%%%%%%%%%%%%%%%%%%%%%%%%%%%%%%%%
where $\lambda$ denotes the oscillation length,
%%%%%%%%%%%%%%%%%%%%%%%%%%%%%%%%%%%%%%%%%
\be
\lambda = 2.47~ 
\frac{\mathrm eV^2}{\Delta m^2} \frac{E}{\mathrm MeV}~ m.
\ee
%%%%%%%%%%%%%%%%%%%%%%%%%%%%%%%%%%%%%%%%%
On the other hand, 
if we neglect the Earth matter 
effects in the solar neutrino transitions,
which are rather small,
the $\nu_e$ survival probability 
relevant for the interpretation of the data of the
SNO and SK solar neutrino experiments,
is given by the adiabatic MSW prediction \cite{msw} 
%%%%%%%%%%%%%%%%%%%%%%%%%%%%%%%%%%%%%
\be 
P_{ee}^{sno}\approx\sin^2\theta_{12}.
\label{P2sno}
\ee
%%%%%%%%%%%%%%%%%%%%%%%%%%%%%%%%%%%%%
Since $\sin^22\theta_{12}$ is a less sensitive function of 
$\theta_{12}$ compared to 
$\sin^2\theta_{12}$, the survival probability 
relevant for the interpretation of the
\kl data is less sensitive to 
$\theta_{12}$ than that measured at SNO. Moreover, 
the average $e^+-$ energy 
measured at \kl ($\sim 5$ MeV),
and the average source-detector distance for \kl
($\sim 180$ km),  correspond to $L \sim \lambda$ 
for the best-fit $\Delta m^2_{21}$. 
At $L \sim \lambda$,
the $\bar{\nu}_e$ survival probability has a maximum (SPMAX).
This means that the coefficient 
of the $\sin^22\theta_{12}$ term 
in $P_{ee}^{KL}$ is relatively small,  
preventing the \kl experiment to reach
high precision in the determination of $\theta_{12}$.
Evidently, the sensitivity to $\theta_{12}$ can be improved 
by reducing the baseline length to 
$L \sim \lambda/2$, corresponding to a minimum of the 
$\bar{\nu}_e$ survival probability (SPMIN) \cite{th12}. 
We shall come back to this point later.  

%%%%%%%%%%%%%%%%%%%%%%%%%%%%%%%%%%%%%%%%%%%

\section{Consistency Check between Different Experiments}

%%%%%%%%%%%%%%%%%%%%%%%%%%%%%%%%%%%%%%%%%%%%%%%%%%%%%%

\begin{figure}[t]
\centerline{\psfig{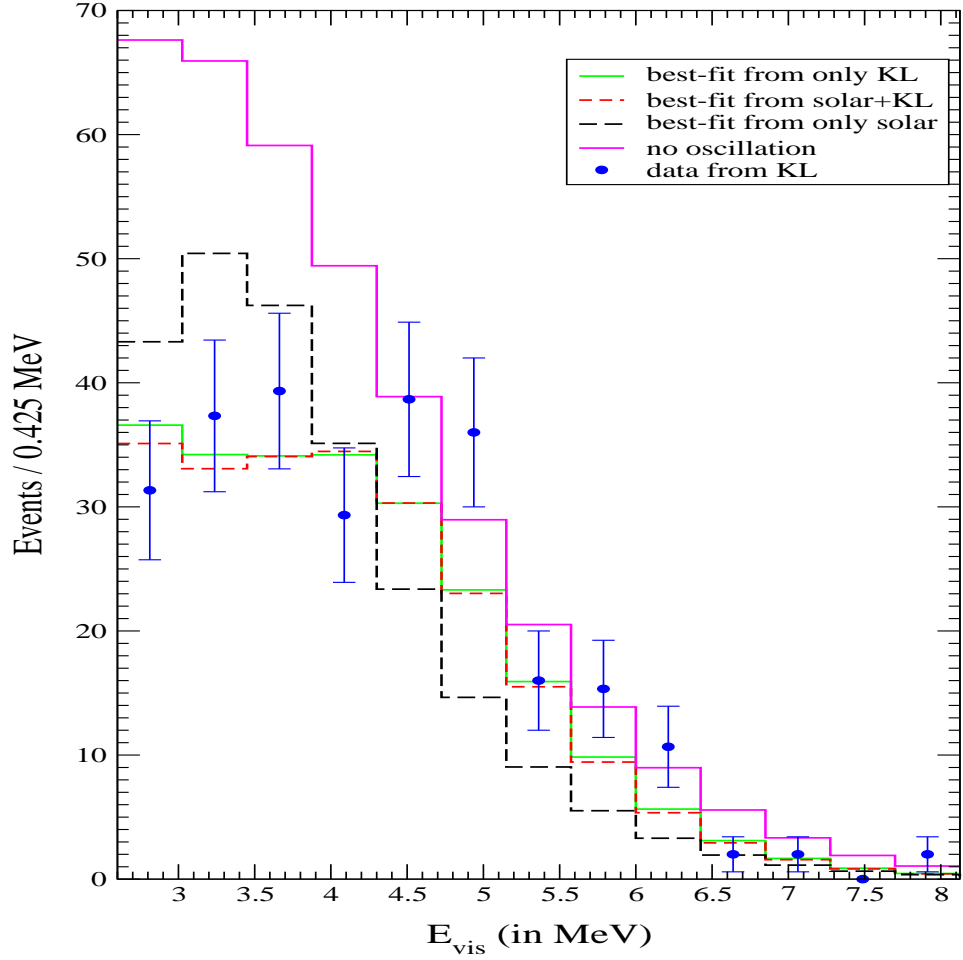}}
\caption{\label{spec}
Predictions for the spectrum measured in the \kl experiment, 
for values of $\Delta m^2_{21}$ and $\sin^2\theta_{12}$
corresponding to the i) solar neutrino data best-fit point, 
ii) \kl spectrum data best-fit point, 
and iii) solar neutrino +\kl spectrum data best-fit point. 
Also shown are the \kl 766.3 Ty spectrum data points
from \cite{kl766}. We also show the unoscillated spectrum 
obtianed by our code.  
}
\end{figure}
%%%%%%%%%%%%%%%%%%%%%%%%%%%%%%%%%%%%%%%%%%%%%%%%%%%%%

In this Section we check the 
consistency between the allowed regions obtained using data 
from different experiments.
In Fig. \ref{spec} we compare the \kl spectrum data with 
the predictions for the spectrum obtained
for values of $\Delta m^2_{21}$ and $\sin^2\theta_{12}$
corresponding to the i) solar neutrino data best-fit point, 
ii) \kl spectrum data best-fit point, 
and iii) solar neutrino +\kl spectrum data best-fit point. 
We also show the unoscillated spectrum obtained by us. 
This agrees fairly well with that given by the \kl collaboration 
in \cite{kl766} indicating that we have correctly 
implemented the reactor power and operation schedules from the 
availbale sources.  
This figure clearly illustrates
the sensitivity of the predicted spectrum to $\ms$, 
and the deviations  of the observed 
spectrum from that predicted 
at the solar neutrino data best-fit point.
%%%%%%%%%%%%%%%%%%%%%%%%%%%%%%%%%%%%%
\begin{table}
\begin{center}
\begin{tabular}{c|c|c|c}
\hline
Experiment & Observed rate/BP04 prediction & Predicted Rate & Predicted Rate \\
& & at global best-fit & at solar best-fit \\
\hline
Ga & $0.52 \pm 0.029$ & 0.546 & 0.538 \cr
Cl & $0.301 \pm 0.027$ & 0.350 & 0.346 \cr
SK(ES) & $0.406 \pm 0.014$ & 0.395 & 0.396 \cr
SNO(CC) & $ 0.274 \pm 0.019$ & 0.291 & 0.290  \cr
SNO(ES) & $ 0.38 \pm 0.052 $ & 0.387 & 0.387 \cr
SNO(NC) & $0.895 \pm 0.08 $ & 0.879 & 0.903 \cr 
\hline
\end{tabular}
\label{rates}
\caption{The observed rates w.r.t predictions from the
latest Standard Solar Model
\cite{bp04}. We also show the predicted rates for 
the best fit values of $\Delta m^2_{21}$ and $\sin^2\theta_{12}$,
obtained in the analysis of
the i) global solar neutrino data, and
ii) global solar neutrino +\kl data.  
}
\end{center}
\end{table}
%%%%%%%%%%%%%%%%%%%%%%%%%%%%%%%
In the absence of the \kl results,
it was necessary to compare the ``low'' and ``high'' energy CC 
solar neutrino data to determine $\Delta m^2_{21}$, 
and to compare the CC and NC data to 
determine $\sin^2\theta_{12}$.
With $\ms$ determined using the new \kl results,
one can make a consistency check 
by dispensing with the data from anyone of the solar neutrino 
experiments. 

\begin{figure}[t]
\centerline{\psfig{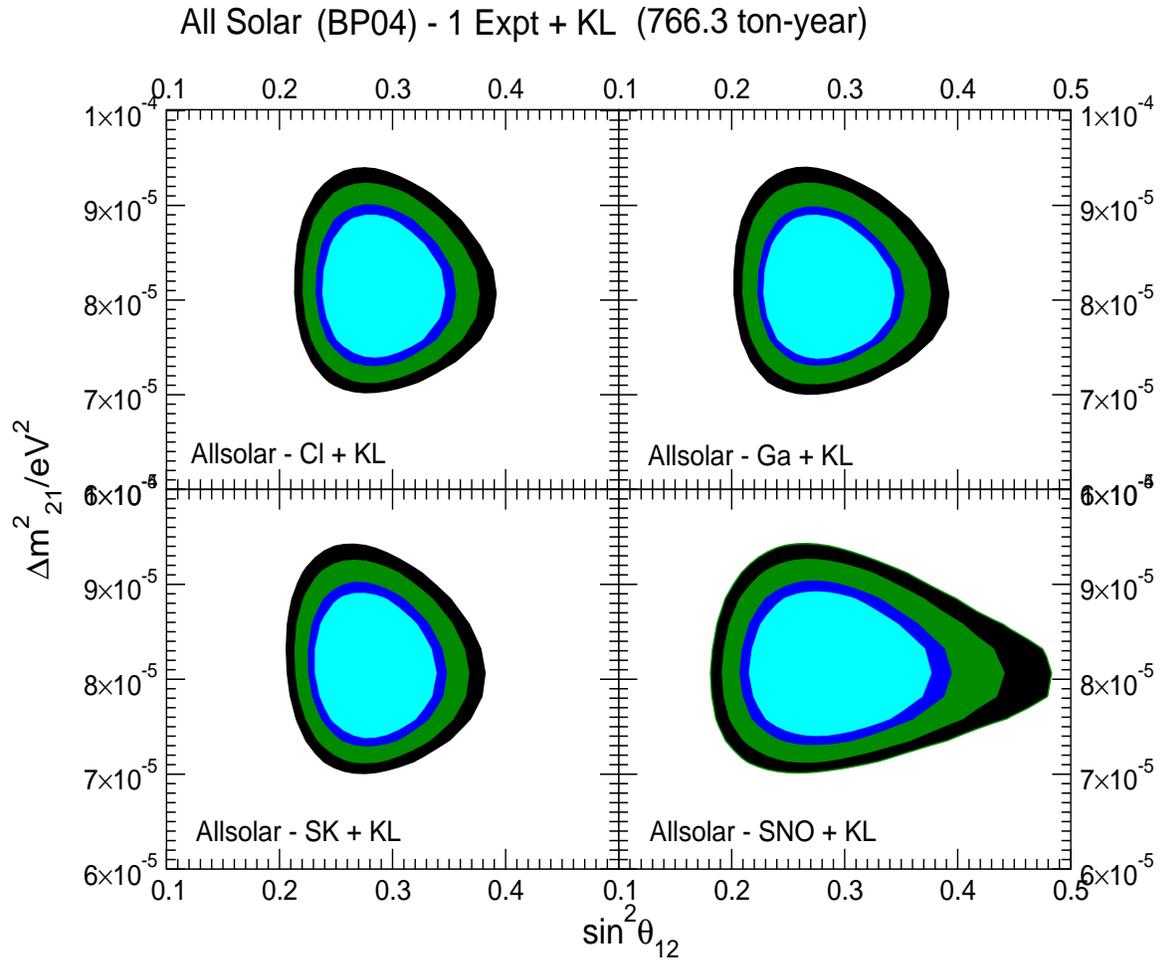}}
\caption{\label{all-1}
The allowed regions in $\dm$ and $\sss$ parameter space,
obtained by removing the data from 
one solar neutrino experiment from the global fit. 
}
\end{figure}

  In Fig. \ref{all-1} we present the allowed 
regions obtained by taking out 
the data from one solar neutrino experiment 
from the global data set. 
As before we let the $^{8}{B}$ 
flux normalisation vary freely in the analysis.  
The figure shows that the allowed 
region is robust and does not change 
considerably
when the data from one experiment is left out of the analysis.
Taking out the SNO data will lead to 
smaller values of $f_B$, 
and correspondingly larger values 
of $\sin^2\theta_{12}$ being allowed.
Using the 162 Ty \kl data and taking out the 
SNO results from the analysis, made 
the maximal mixing solution allowed
\cite{sgnu04}. With the 766.3 Ty \kl data 
included in the analysis, the
maximal mixing is ruled out 
at 3$\sigma$ even leaving out the SNO data from 
the data set used in the analysis. 
This is a consequence  of the increased  
precision of the \kl data
which 
%strongly 
disfavours 
the maximal mixing solution. 
%%%%%%%%%%%%%%%%%%%%%%%%%%%%%%%%%%
%%%%%%%%%%%%%%%%%%%%%%%%%%%%%%%%%%%
%
In Table 2 we compare the observed 
event rates in the different solar 
neutrino experiments with those predicted
for the best fit values of 
$\dm$ and $\sss$, obtained in the
analysis of the global solar neutrino data
and global solar neutrino + \kl spectrum data. 
Note that the SNO NC event rate relative 
to the SSM prediction of BP04 is $f_B = 0.895 \pm 0.08$,
while its earlier central value, with respect to the 
SSM prediction of BP00, was slightly above 1. 
This drop simply reflects the increase in the central 
value of the $^{8}{B}$ flux from $5.1  \times 10^{-5}$ cm$^{-2}s^{-1}$ 
to $5.8 \times 10^{-5}$ cm$^{-2}s^{-1}$ 
in the latest SSM of BP04 \cite{bp04}. 
There is a corresponding drop 
in the other experimental rates shown in Table 2. However, this 
renormalisation has no effect on our results since we have not used 
SSM prediction for the 
normalisation of the $^8B$ flux. Instead we have left $f_B$ 
as a free parameter to be determined by the solar neutrino data. 
This parameter is primarily determined by the NC event rate 
measured in SNO. 
We see from Table 2 that all the measured event rates, 
except that observed in the Cl experiment, 
are in very good agreement with the predicted ones. 
There is very little difference 
between the predictions, corresponding to the 
solar neutrino data and solar neutrino + \kl spectrum data
best-fit points. 
This shows the insensitivity of the fit of the
solar neutrino data to small 
variations in $\Delta m^2$, in contrast 
to the fit of the \kl spectrum data. 
%Only the Cl experiment event rate shows
%a modest increase due to the 
%upward shift of the 
%$\Delta m^2$ best fit value of 
%the solar neutrino + \kl spectrum data
%with respect to the  best fit value
%of the solar neutrino data.
Note that the obtained Cl rate is 
by 2$\sigma$ lower than the global
best-fit prediction. This is 
a statistically small but well known deviation 
which cannot be explained by the LMA solution \cite{eindep}. 
If such a deviation is confirmed by future intermediate energy solar 
neutrino experiments like Borexino,
it will call for some 
additional subdominant mechanism 
of solar neutrino transitions.

%%%%%%%%%%%%%%%%%%%%%%%%%%%%%%%%%%%%%%%%%%%%%%%%%%%%%%%%%%%
%\vspace{-0.4cm}
\section{Three Flavour Neutrino Mixing Analysis}
%\vspace{-0.3cm}
%%%%%%%%%%%%%%%%%%%%%%%%%%%%%%%%%%%%%%%%%%%%%%%%%%%%%%%%%%%

%%%%%%%%%%%%%%%%%%%%%%%%%%%%%%%%%%%%%%%%%%%%%%%%%%%
\begin{figure}[t]
\centerline{\psfig{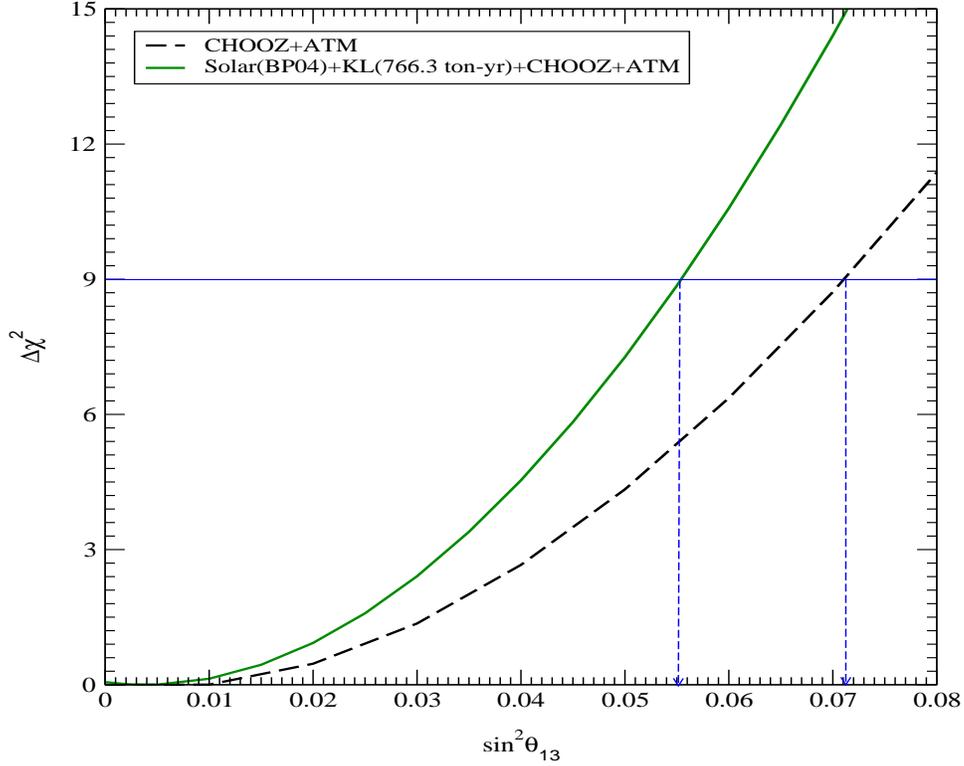}}
\caption{\label{delchith13}
Bounds on the mixing angle $\theta_{13}$ from
the CHOOZ data only (dashed line)
and the combined solar, CHOOZ and \kl 
data (solid line).
The $\Delta m^2_{31}$ is allowed to vary within the 3$\sigma$ range 
(($1.28-4.17) \times 10^{-3}$ eV$^2$),
obtained using 
the one parameter $\Delta \chi^2$ vs $\Delta m_{31}^2$ fit 
of the SK atmospheric Zenith angle data, 
presented by the SK Collaboration in Neutrino 2004 \cite{kearns04}. 
All the other parameters are allowed to vary freely.
The short-dashed lines show the $3\sigma$ limits 
corresponding to the case of
1 parameter fit.
}
\end{figure}
%%%%%%%%%%%%%%%%%%%%%%%%%%%%%%%%%%%%%%%%%%
%

In this Section we present results obtained from the analysis 
of the global data on solar and reactor neutrinos within a 
three-flavor neutrino mixing framework. In this case 
$\Delta m^2_{21} \cong  \Delta m^2_{\odot}$ and 
$\Delta m^2_{31} \cong  \Delta m^2_{atm}$. 
The  best-fit value of $\Delta m^2_{atm}$,
obtained in the latest two-neutrino mixing 
analysis of the Super-Kamiokande 
atmospheric neutrino data 
on the Zenith-angle 
distribution of the $\mu-$like events 
is 2.1$\times 10^{-3}$ eV$^2$ \cite{kearns04}. 
Thus, the two-neutrino mixing
analyses of the solar and atmospheric neutrino data
indicate that $\Delta m^2_{21} << \Delta m^2_{31}$.
Under this approximation 
the effect of the third 
heaviest neutrino in the relevant 
solar neutrino and reactor anti-neutrino
survival probabilities 
is due mainly to
the mixing angle $\theta_{13}$. 
The relevant $\nu_e$ and $\bar{\nu}_e$ 
survival probabilities in the three-neutrino mixing
cases of interest are given by the following expression:
%%%%%%%%%%%%%%%%%%%%%%%%%%%%%%%%%%%%%
\be
P_{ee}^{3\nu} \cong \cos^4 \theta_{13} P_{ee}^{2\nu} + \sin^4\theta_{13},
\label{3genpee}
\ee
%%%%%%%%%%%%%%%%%%%%%%%%%%%%%%%%%%%%
%
where $ P_{ee}^{2\nu}$ is the $\nu_e$ or
$\bar{\nu}_e$
survival probability in the case of
two-neutrino mixing (see, e.g., \cite{SP3nuosc88}). 
For solar neutrinos,
$ P_{ee}^{2\nu} \equiv P_{ee\odot}^{2\nu}$ 
is the two-neutrino mixing
$\nu_e$ survival probability  \cite{SP88} with
the solar electron number density $N_e$
replaced by $N_e \cos^2\theta_{13}$.
In the case of \kl experiment one has
$ P_{ee}^{2\nu} \equiv P_{eeKL}^{2\nu}$, where
$P_{eeKL}^{2\nu}$ is given by eq. (\ref{probkl})
in which $\Delta m^2 \equiv \Delta m^2_{21}$.

 Strong constraints on the value of $\theta_{13}$ 
have been obtained in the
CHOOZ and Palo Verde reactor 
antineutrino experiments \cite{chooz}. 
We include the CHOOZ
results in our three-flavour neutrino mixing
analysis (see also \cite{BNPChooz}).  
In the limit of $\Delta m^2_{21} << \Delta m^2_{31}$,
the probability relevant for the interpretation of the 
CHOOZ data is given by 
%%%%%%%%%%%%%%%%%%%%%%%%%%%%%
\be
P_{eeCHOOZ}^{3\nu} \cong 1 - \sin^22\theta_{13}
\sin^2(\Delta m_{31}^2 L/4E).
\ee 
%%%%%%%%%%%%%%%%%%%%%%%%%%%%%%%%%
We note that 
$P_{eeCHOOZ}^{3\nu}$ 
depends on $\Delta m^2_{31}$, 
while $P_{ee\odot}^{2\nu}$ 
and $P_{eeKL}^{2\nu}$ depend 
on $\Delta m^2_{21}$.

   We allow $\Delta m^2_{31}$ to vary freely within  the 
$3\sigma$ range 
($1.28-4.17) \times 10^{-3}$ eV$^2$,
obtained using  the one parameter 
$\Delta \chi^2$ vs $\Delta m_{31}^2$ fit 
of the SK atmospheric neutrino 
Zenith angle data, 
presented by the SK Collaboration at the Neutrino 2004
International Conference \cite{kearns04}
\footnote{The allowed range of $\theta_{13}$ depends crucially 
on the allowed range of the atmospheric mass squared difference 
$\Delta m^2_{31}$ \cite{sgnu04}.},
and perform a combined three-neutrino oscillation 
analysis of the global solar 
neutrino and reactor anti-neutrino
data, including both the \kl and CHOOZ results.   
The best-fit values of the parameters 
obtained from the three-flavor neutrino
mixing analysis are: 
%%%%%%%%%%%%%%%%%%%%%%%%%%%%%%%%%
\begin{equation}
\ms = 8.0\times 10^{-5}~{\rm eV^2},~~\sss = 0.28,~~ 
\sin^2\theta_{13}=0.004,~~f_B=0.88, ~~\chi^2_{min}/d.o.f. = 91.68/105 \qquad 
\end{equation}
%%%%%%%%%%%%%%%%%%%%%%%%%%%%%%
%
%%%%%%%%%%%%%%%%%%%%%%%%%%%%%%%%%%%%%%%%
\begin{figure}
\centerline{\psfig{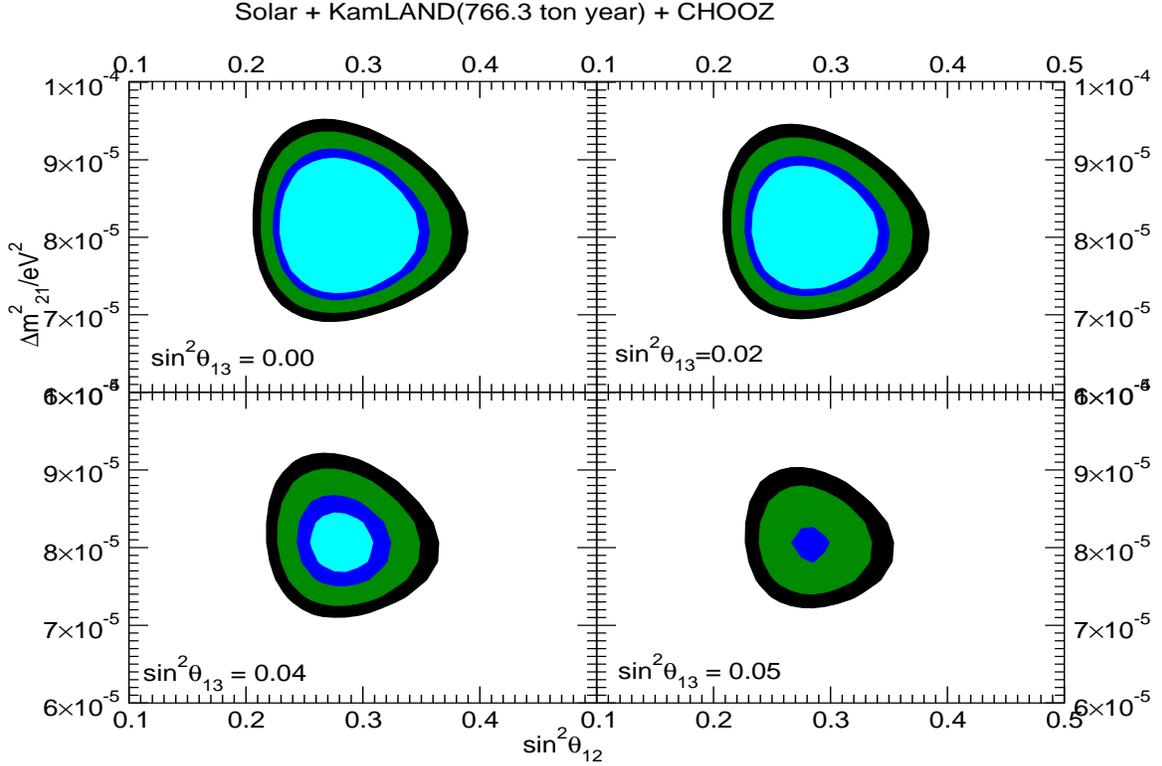}}
\caption{\label{solkl3osc}
The 90\%, 95\%, 99\% and 99.73\% C.L. allowed 
regions in the $\dm-\sss$ plane, obtained 
in a three-neutrino oscillation analysis of the 
global solar and reactor neutrino data, 
including the data from the 
\kl and CHOOZ experiments. The different panels 
correspond to different fixed values of $\sin^2\theta_{13}$. 
In the analysis $\Delta m^2_{31}$ 
is allowed to vary freely within $(1.28-4.17)\times 10^{-3}$ eV$^2$
taken from \cite{kearns04}.
Here we use three parameter fit 
$\Delta \chi^2$ values to plot the C.L. contours.
}
\end{figure}
%%%%%%%%%%%%%%%%%%%%%%%%%%%%%%%%%%%%%%%%%%%%%

   In Fig. \ref{delchith13} we present the $\Delta \chi^2$ 
as a function of $\sin^2\theta_{13}$, for  
$\Delta m^2_{31}$ allowed to vary 
within its 3$\sigma$ allowed range \cite{kearns04},
$(1.28-4.17)\times 10^{-3}$ eV$^2$,
and the other parameters allowed to vary freely. 
The $3\sigma$ bounds on 
$\sin^2\theta_{13}$, obtained from CHOOZ data analysis,
can be directly read from the 
figure for a one parameter fit as 
$\sin^2\theta_{13} < 0.07$. 
The bound  derived from the 
combined analysis of the solar neutrino, 
CHOOZ and KamLAND data
is $\sin^2\theta_{13} < 0.055$.

  In Fig. \ref{solkl3osc} we show the allowed regions 
in the $\dm-\sss$ plane for 
four fixed values of $\theta_{13}$. 
We note that although 
the presence of a small non-zero $\theta_{13}$ can  
improve the fit in the regions of 
the parameter space with higher values of $\dm$  \cite{snosaltus},
i.e., in the high-LMA zone, the new \kl data are able to exclude the 
high-LMA region at more than 3$\sigma$ even in the presence of a 
third generation in the mixing, 
indicating the robustness of the low-LMA solution. 

%%%%%%%%%%%%%%%%%%%%%%%%%%%%
\vspace{-0.4cm}
\section{Future Projections}
\vspace{-0.3cm}
%%%%%%%%%%%%%%%%%%%%%%%%%%%%

\begin{figure}[t]
\centerline{\psfig{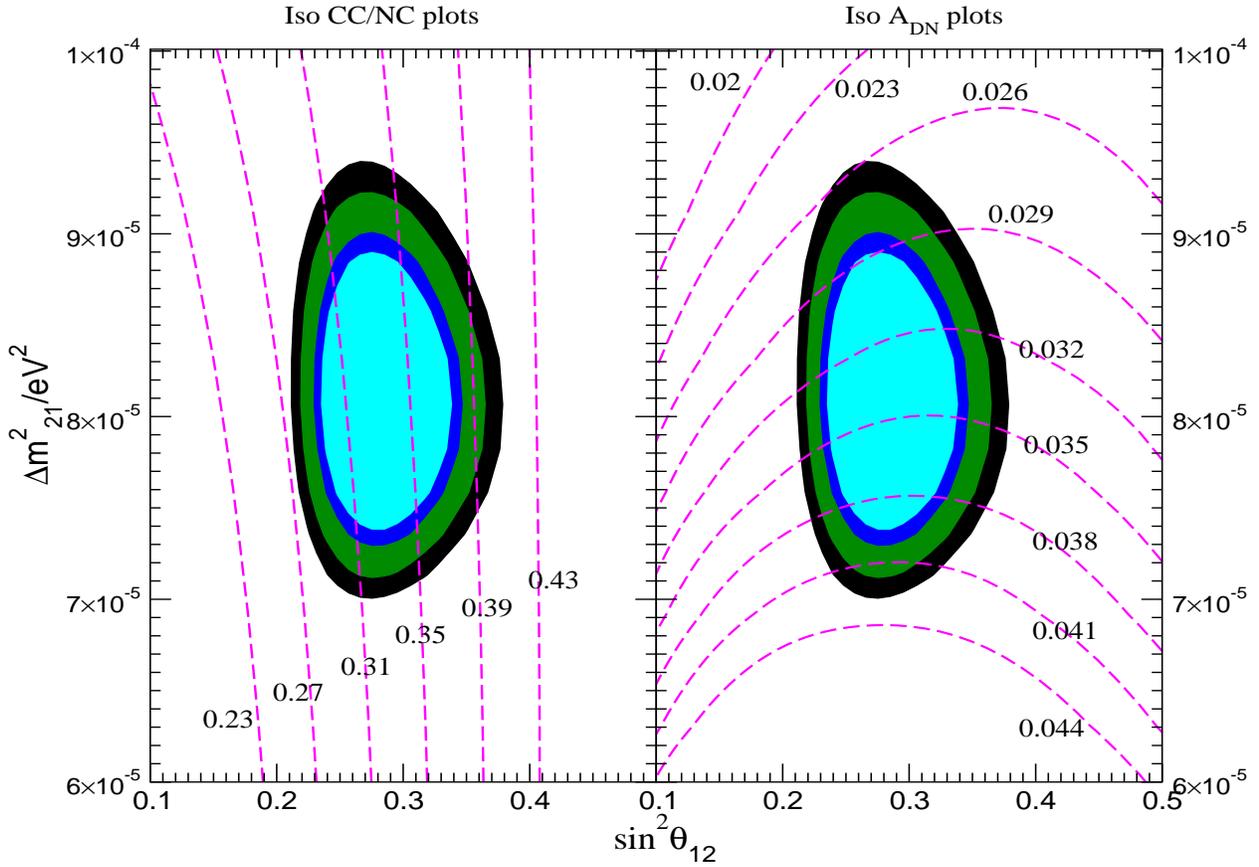}}
\caption{\label{dniso}
Lines of constant day-night asymmetry (right panel) and 
iso-CC/NC contours (left panel) for the SNO
experiment, superposed on the allowed region obtained 
in the global 
analysis of the solar neutrino and \kl data.} 
\end{figure}
%%%%%%%%%%%%%%%%%%%%%%%%%%%%%%%%%%%%%

The recent \kl data combined with the
solar neutrino data unambiguously 
determine the low-LMA solution as unique
solution of the solar neutrino problem. 
It also enables us to determine $\dm$ with a relatively
high precision:  
$\sim$ 10\% at 90\% C.L.
The high-LMA solution  is disfavored 
at more than 3$\sigma$. 
The uncertainty in the value of $\ms$ is expected to 
diminish further as \kl collects more data.
However, as we have stressed before, the \kl 
experiment does not appreciably 
reduce the error on  the value of $\sss$ \cite{th12}.

  In the near future, the SNO collaboration
is expected to publish data on the CC ($e^-$) 
day/night spectrum, observed during the salt phase of 
the experiment. The recent \kl results allow us
to make relatively precise predictions 
for the the day-night asymmetry 
in the SNO experiment: 
%%%%%%%%%%%%%%%%%%%%%%%%%%
\be
A_{DN} = 2\frac{N-D}{N+D}.
\ee
%%%%%%%%%%%%%%%%%%%%%%%%%%%
%
In the right hand panel of Fig. \ref{dniso} 
we show the lines 
of constant $A_{DN}$ values in the 
$\dm-\sss$ plane for the SNO experiment
(see also, e.g., \cite{maris1,maris2}).
The predicted $A_{DN}$ in SNO  
for the current best-fit values of the parameters
and the corresponding
$3\sigma$ range, are given by:
%%%%%%%%%%%%%%%%%%%%%%%%%%%%%%%%%
\be
A_{DN}(SNO) = 0.034,~~3\sigma~ {\rm range}:~0.027-0.043.
\ee
%%%%%%%%%%%%%%%%%%%%%%%%%%%%
The published SNO result on the  D/N asymmetry is 
$A^{exp}_{DN}(SNO) = 7 \pm 5\%$. 
Thus, the error has to be reduced to $\sim 1\%$   
in order to observe $A_{DN}(SNO) \neq 0$
at $\sim$ 3$\sigma$ level.

 In the left hand panel of the same  
figure we also plot the iso-CC/NC contours for SNO. 
The measured value of the CC to NC ratio from the salt phase 
of SNO experiment is,
%%%%%%%%%%%%%%%%%%%%%%%%%%%%
\be
R_{CC/NC}^{salt}=0.305 \pm 0.033.
\ee
%%%%%%%%%%%%%%%%%%%%%%%%%%%
%
The phase III of SNO will collect 
neutral current data using Helium counters \cite{sno3}. This 
would give a totally uncorrelated information on the CC and NC 
event rates observed in SNO and a reduced error 
for the NC event rate.
The projected total error for the observed NC event rate for  
this phase is 6\% \cite{sno3}. 
For the CC event rate we assume that the 
statistical error during the 
phase III would be approximately
the same as in each of the earlier two phases, 
while the systematic error is taken
to be 4.5\%, i.e., slightly smaller than
the 5\% reported in phases I and II of SNO.  
Thus, we assume that the total error in CC event rate 
measurement from all the three phases 
combined will be about 5\%. If 
the central value of the CC and NC event rate
ratio would remain the same as observed in the salt phase,
the CC/NC ratio expected to be measured 
in the phase III of the SNO 
experiment would be
%%%%%%%%%%%%%%%%%%%%%%%%%%%
\be
R_{CC/NC}^{He}=0.305 \pm 0.024.
\ee
%%%%%%%%%%%%%%%%%
Since for $\dm \ltap 10^{-4}~{\rm eV^2}$
the CC/NC ratio in SNO is mainly related to the solar 
neutrino mixing angle (see, e.g., ref. \cite{maris2},
eq. (\ref{P2sno}) and the left hand panel in 
Fig. \ref{dniso}),
the reduction in the $R_{CC/NC}$ error 
is expected to result in 
an improvement in the precision of $\sss$
determination. 
We have made a projected analysis 
of the global solar neutrino data, 
including the upgraded CC and NC errors expected from the 
phase III of the SNO experiment. The 
range of allowed values of $\sss$ could be 
reduced to about $0.22-0.34 ~(0.21-0.36)$ 
at 99\% C.L. (99.73\% C.L.), 
corresponding to a
spread of $21\%~(26\%)$  \cite{skgd}. 

   There has been a recent proposal of 
adding 0.1\% gadolinium to the water in 
the Super-Kamiokande detector to improve 
the detector sensitivity to neutrons \cite{gadzooks}. 
This would result in a remarkable increase 
in the detector sensitivity to 
low energy $\anue$, transforming SK into a huge reactor 
antineutrino detector (SK-Gd), 
with an event rate that is about 43 times 
larger than that observed in KamLAND 
\cite{gadzooks,skgd}. After 5 years of data taking, the
SK-Gd experiment can measure $\ms$ with 
$\sim 1\%$ and $\sss$ with $\sim 15\%$ uncertainty at 99\% C.L.
\cite{skgd}.

   As discussed earlier, a very precise measurement of $\sss$ 
can be achieved in a reactor experiment
with a baseline corresponding to an SPMIN 
of the $\bar{\nu}_e$ survival probability
\cite{th12}. The condition for SPMIN is
$L \cong \lambda/2 = 1.24  (\mathrm E/MeV)
(\mathrm eV^2/\Delta m^2_{12})$ m.
For the low-LMA solution region 
and the average energy of the $e^+$ observed in the
\kl experiment, 
this corresponds to a 
distance of approximately (50 - 70) km \cite{th12}. 
For an experiment with a 70 km baseline and 24.3 GW 
(18.6 GW) reactor power, corresponding to the Kashiwazaki
(Daya Bay) complex in Japan (China),
$\sss$ can be determined with a  
$\sim 10\%$ error at 99\% C.L. with a 3 kTy (4 kTy) 
statistics \cite{th12}.

    The forthcoming solar neutrino experiments are  
Borexino \cite{borex} and KamLAND-$^7Be$,
which will provide an accurate 
measurement of the $^7Be$ neutrino flux, 
and the Low energy solar Neutrino (LowNu) experiments
\cite{XMASS,LENS}, which are designed to
measure the flux of solar $pp$ neutrinos.
The potential of Borexino and any generic LowNu
experiment \cite{lownu} in constraining the 
solar neutrino oscillation 
parameters have been studied recently in 
\cite{th12,Bahcall:2003ce}.
For the allowed regions obtained in this paper, 
we find the predicted rates
for Borexino and LowNu experiments to be 
%%%%%%%%%%%%%%%%%%%
\be
R_{Be} &=& 0.67,~~(3\sigma~ {\rm range}:~ 0.62-0.72)\\
R_{pp} &=& 0.71,~~(3\sigma ~{\rm range}:~ 0.67-0.76).
\ee
%%%%%%%%%%%%%%%%%%%%%%%

%%%%%%%%%%%%%%%%%%%%
\vspace{-0.4cm}
\section{Conclusions} 
\vspace{-0.3cm}
%%%%%%%%%%%%%%%%%%%%%

We have investigated the 
implications of including the
recent \kl spectrum data 
in global solar neutrino 
oscillation analysis. 
The observed spectral distortion in the 
\kl experiment firmly establishes 
\dm to lie in the low-LMA solution region.
The high-LMA solution is 
excluded at more than  4$\sigma$ 
by the global solar neutrino and \kl spectrum data.
The maximal solar neutrino mixing is 
ruled out at $6\sigma$ level.
We have found that the $3\sigma$ allowed region in the 
$\ms-\sss$ plane remains remarkably stable even when we 
leave out the data from one 
of the solar neutrino experiments from the global fit.
Likewise, there is practically no increase 
in the allowed region when one goes from two to three flavor
neutrino oscillation analysis of the global solar neutrino 
and \kl spectrum data. The $3\sigma$ upper limit on $\sin^2\theta_{13}$ 
was found to be $ \sin^2\theta_{13} <0.055$.
We have derived predictions 
for the CC to NC event rate ratio
and day-night (D-N) asymmetry in the CC event rate,
measured in the SNO experiment, and for the 
suppression of the event rate in the BOREXINO 
and LowNu experiments, designed to measure the
$^7Be$ and $pp$ solar neutrino fluxes.
With the value of $\dm$ 
determined more precisely  
using the current solar neutrino and \kl data,
the predicted range of possible 
values of the
day-night asymmetry in 
the CC event rate at SNO 
narrows down to (0.025 - 0.041)
at 99.73\% C.L.
Remarkably high precision in the measurement
of $\dm$ can be achieved 
with the Super-Kamiokande detector
loaded with 1\% gadolinium: this  
would transform Super-Kamiokande
into a huge reactor $\bar{\nu}_e$
detector (SK-Gd) with an event rate 43 times
larger than that observed in the \kl
experiment. Finally, we have discussed  
how the precision of $\sin^2\theta_{12}$
determination can improve 
with the increasing of the precision of the
future SNO data, by the SK-Gd reactor 
$\bar{\nu}_e$ oscillation experiment, 
as well as, by performing a reactor 
$\bar{\nu}_e$ oscillation experiment with a baseline
of $\sim 70$ km.

    With the publication of the latest \kl data
the neutrino oscillation origin of 
the observed solar neutrino deficit
is firmly established. The future 
high precision measurements of the solar neutrino
oscillation parameters will be 
of fundamental importance for 
understanding the true origin 
of the flavour neutrino mixing.

\vskip 8pt
S.G. and D.P.R. would like to thank 
respectively SISSA and 
The Abdus Salam International Centre for
Theoretical Physics for hospitality.
This work was supported by  
the Italian INFN under the 
program ``Fisica Astroparticellare'' (S.T.P.).

\vspace{-0.3cm}
%%%%%%%%%%%%%%%%%%%%%%%%%%%%

%%%%%%%%%%%%%%%%%%%%%%%%%%%%

\end{document}